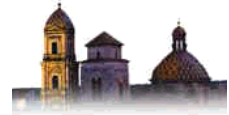

# In the debris of hadron interactions lies the beauty of QCD  (part II)

M.R. Pennington*

Institute for Particle Phenomenology, University of Durham, Durham DH1 3LE, U.K.

Recent progress in understanding the strong physics regime of QCD is described. The role played by condensates, particularly $\langle q\overline{q}\rangle$, in breaking chiral symmetry and generating constituent masses for $u$ and $d$ quarks is reviewed. The influence this has on hadrons with vacuum quantum numbers is emphasised. What we know of this sector from recent experiments on $\phi$-radiative decays and from $D$ decays to light hadrons is discussed. How we may gain a more complete understanding of this vacuum sector is outlined.

## 1  Structure of the QCD vacuum

To learn about the underlying theory of quark and gluon interactions we have to study hadronic collisions. In the debris of these lies the beauty of QCD. The task required to reveal this world of quarks and gluons is akin to an archaeological dig through the debris of Iraq in 2003 to unearth the civilisation of Babylon beneath.

Over 30 years of digging considerable progress has been made in uncovering the strong physics aspects of QCD. What makes QCD more interesting than QED is the nature of the vacuum. For QED the vacuum is described by perturbation theory. It is essentially empty with a low density of particle-antiparticle pairs. In QCD because the interactions are stronger not only is the vacuum a denser sea of $q\overline{q}$ pairs and a cloud of gluons, but so strong are the forces that condensates of quarks, antiquarks and gluons form in all colour singlet combinations. That of lowest dimension is the $q\overline{q}$ condensate. The scale of this condensate characterises the key non-perturbative effects of light hadron systems. Being non-zero it dynamically breaks the chiral symmetry of QCD. This simultaneously ensures that the pion is the Goldstone boson of this symmetry breaking and that the corresponding scalar field is the Higgs sector of the strong interaction, responsible for the masses of all light hadrons.

The value of this condensate $\langle u\overline{u}\rangle \simeq \langle d\overline{d}\rangle$ can be found in at least three different ways [1,2], which here will be described as "phenomenological", "experimental" and "theoretical". Remarkably the 3 distinct ways give consistent results. The first and oldest phenomenological method is the application of the QCD sum-rules of Shifman, Vainshtein and Zakharov [3] to scalar and pseudoscalar currents, a subject to which the Bari group [4] have made important contributions. The sum-rules relate the matrix element of current correlators evaluated at low energies from hadronic data to their calculation at higher energies using the Operator Product Expansion. It is in this expansion that condensates arise. Though such sum-rules have been studied for 25 years, recent precision has come from a better understanding of the use of contour improvement, of pinched weights in finite energy sum-rules and technological advances in the calculation of higher order corrections in perturbative QCD (see citations in Ref. 2). Agreement between the theoretical and experimental sides of the sum-rules gives $\langle q\overline{q}\rangle \simeq -(250 \pm 25 \text{ MeV})^3$ at a scale of 2 GeV. However, sum-rule analyses require an artistry that makes them at best a consistency check on the size of condensates, not an absolute determination.

Much more direct experimental confirmation is obtained by measuring low energy $\pi\pi$ scattering precisely [1,5–7]. To see how, first consider the process with one of the initial or final pions massless. Then in the limit of zero momentum for this pion, the amplitude vanishes, as Adler [8] first deduced. Thus at the symmetry point in the middle of the Mandelstam triangle at $s = t = u = m_\pi^2$, the amplitude for this strong interaction process is zero. The interaction will continue to be *weak* in the neighbourhood of this point. As a consequence the amplitude can usefully be expanded as a Taylor series in powers of $s, t$ and the square of the mass of the off-shell pion. The scale of this expansion is naturally given by a typical meson scale of $m_\rho^2$, or equally $32\pi f_\pi^2$ in the chiral expansion [9]. Pions being the Goldstone bosons of chiral symmetry breaking know about the size of $\langle q\overline{q}\rangle$. It is this that determines the position of the on-shell appearance of the Adler zero and is reflected in the value of the $\pi\pi$ scattering amplitude at the on-shell symmetry point $s = t = u = 4m_\pi^2/3$. As the $q\overline{q}$ condensate decreases in scale from 250 MeV, the position of the Adler zero within the Mandelstam triangle moves away from the symmetry point, making the amplitude there increase by up to a factor of 4 [6]. Now the zero of an analytic function of several complex variables lies within a surface of zeros. At each energy, $\sqrt{s}$, this surface produces a dip at some scattering angle (or value of $t$) in the cross-section for $\pi^+\pi^0 \to \pi^+\pi^0$, for instance. As the energy increases, this dip, which is observed in experiment, follows a contour traversing the Mandelstam plane. The exact position of this contour in the near threshold region is determined by the size of the $q\bar{q}$ condensate. Precision measurements of $\pi\pi$ interactions below c.m. energies of 450 MeV have now become possible. Combining such data with dispersion relations that incorporate the important constraint of the 3-channel crossing symmetry of $\pi\pi$ scattering allows the position of this zero contour at very low energies to be located, fixing the value of the amplitude at the symmetry point and determining the size of the condensate [7]. While we await the accurate measurement of the amplitude exactly at threshold deduced from the lifetime of pionium [10], one can use the difference of the phase of $S$ and $P$-wave $\pi\pi$ interactions as measured in $K \to e\nu\pi\pi$ decay, as described in detail in Ref. 5. Results from the recent BNL-E852 experiment [11] yield a condensate of $\sim -(270\ \text{MeV})^3$ at 2 GeV, showing that more than 90% of the Gell-Mann-Oakes-Renner relation for $m_\pi^2 f_\pi^2$ expanded in powers of the current quark mass is given by just the first term linear in $m_q$ [12].

---
*An earlier talk with the same title, which is implicitly Part I, with similar topics but a different emphasis will appear in the Proceedings of the Workshop on Gluonic Excitations, held at Jefferson Laboratory, May 2003.

This is to be compared with a "theoretical" determination deduced from the strong limit of QCD in the continuum. This requires solving the Schwinger-Dyson equations under certain plausible assumptions (discussed in Refs. 13,1,2), which enables the behaviour of gluon, ghost and quark propagators and their interactions to be investigated in the small quark mass (or chiral) limit, Fig. 1. Remarkable progress has been made in the past decade [14,13,15]. We understand that if the effective quark-gluon coupling becomes of order unity for momenta below 500 MeV or so (Fig. 1), chiral symmetry is broken. Then a massless current quark (or one of mass of a few MeV), that propagates almost freely over very short distances, has an effective mass of 350 MeV at distance scales of 1 fm. The strong QCD dressing of the $u$ and $d$ quark propagators turns a current quark into a constituent quark. Though such calculations are most conveniently performed in the Landau gauge, this behaviour of the quark propagator can be related [16] to the chiral limit of $\langle q\bar{q}\rangle$, which is gauge invariant. Remarkably, this condensate has a scale of 250-300 MeV [17,15,16], in reassuring consistency with the phenomenology just discussed.

As emphasised by Roberts and collaborators [18], the axial Ward identity ensures that the $q\bar{q}$ bound state with pseudoscalar quantum numbers is the Goldstone boson with its interactions governed by PCAC. In contrast the bound states with scalar and vector quantum numbers have masses reflecting the mass of the fully dressed (or constituent) quark. The behaviour of the gluon and ghost propagators built into these calculations can be compared with Monte Carlo lattice simulations and are in excellent agreement [19,14]. While lattice calculations can only be performed with sizeable quark masses, the continuum Schwinger-Dyson/Bethe-Salpeter system can be computed even in the massless limit with all the essential physics of chiral logs built in. Consequently, this system provides a modelling of the chiral extrapolation [20] so necessary to obtain physically meaningful results for light hadrons on the lattice. The successes described in Refs. 13-15 of this approach to strong physics justify the assumptions needed to truncate the Schwinger-Dyson equations and illustrate how considerable progress has been made in extending the calculability of QCD from the perturbative regime to confinement scales so crucial for light hadron phenomena.

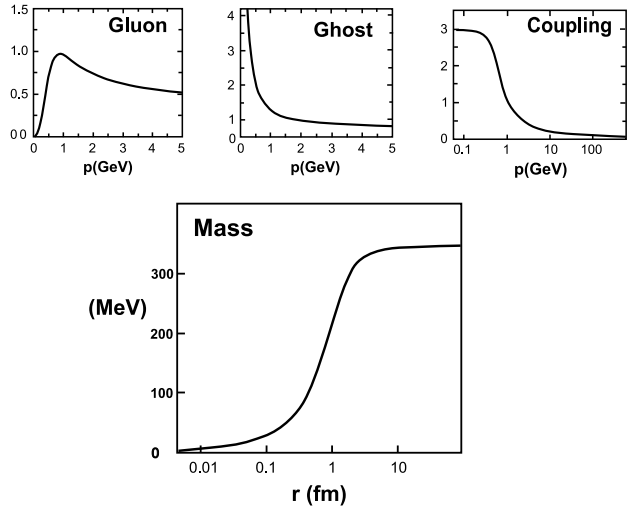

**Figure 1**. Momentum dependence of the Landau gauge gluon and ghost dressing functions from Schwinger-Dyson studies [14]. These give the *effective* quark-gluon coupling shown. How the resulting $u, d$ quark mass function varies with distance of propagation is illustrated.

## 2   Hadrons with vacuum quantum numbers

So far we have learnt that the dynamical breakdown of the chiral symmetry of QCD generated predominantly by $u\bar{u}$, $d\bar{d}$ condensates ensures that pions are Goldstone bosons and the corresponding scalar field plays the role of the Higgs of the strongly interacting sector, its mass and that of all light hadrons reflecting the constituent mass of $u$ and $d$ quarks. But what is this scalar field? Is it the $f_0(400-1200)$ (or $\sigma$), or $f_0(980)$, or $f_0(1370)$, or $f_0(1510)$, or $f_0(1720)$, or some mixture of all of these? None of these states is likely to be a pure $q\bar{q}$ state, none likely to be pure glue, none solely $qq\bar{q}\bar{q}$ or a $K\bar{K}$ molecule. All are mixtures of these, but what mixtures? This is the outstanding issue on which we try to shed a little light.

The hadron states we know best, those that live the longest, are believed to be simply bound states of quarks. States of the quark model are most easily identified with the hadrons we observe experimentally when *unquenching* is unimportant, Fig. 2. For these the theorists' favourite tool, the $1/N_c$ expansion, works. In the large $N_c$ limit, states are stable. Nevertheless the spectrum is very close to that observed. Thus the $\phi$ is readily seen to be an $s\bar{s}$ state and the $\rho$ and $\omega$ combinations of $u\bar{u}$ and $d\bar{d}$. This follows from their respective decays to $K\bar{K}$, and to $2\pi$ and $3\pi$.

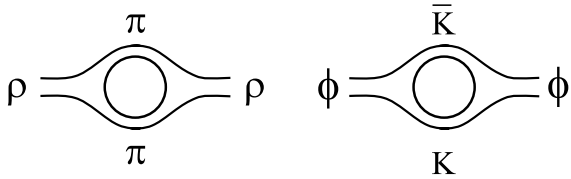

**Figure 2**. "Unquenching" of quark model states to make real hadrons has little effect on the vector mesons, $\rho$ and $\phi$, beyond allowing them to decay.

Though these decay modes are a crucial characteristic of their make-up, they have a relatively small effect on the states themselves. Thus only a small part of the Fock space decomposition of the physical $\phi$ is $K\bar{K}$ : it is predominantly $s\bar{s}$. This is in part because of the $P$-wave nature of its hadronic *dressing*. The resulting small effect reproduces the suppression of the $1/N_c$ expansion.

In contrast, scalar mesons are strongly disturbed by their couplings to open hadron channels [21]. Thus almost regardless of whether their composition is $q\bar{q}$ or $qq\bar{q}\bar{q}$ in the *quenched* approximation, the $f_0(980)$ and $a_0(980)$ are intimately tied to the opening of the $K\bar{K}$ threshold, Fig. 3. Scalars change on unquenching. Their spectrum in the large $N_c$ limit is quite different. For them the $1/N_c$ suppression of quark loops does not occur. The fact that these resonances, $f_0(980)$ and $a_0(980)$, couple to both $\pi\pi/\pi\eta$ and $K\bar{K}$, means scalar non-strange and $s\bar{s}$ states communicate, Fig. 3. The coupling of different flavour quark pairs is not merely unsuppressed, nullifying the OZI rule in the scalar sector [22], but is even enhanced. This reflects the fact that strange quark pairs must exist in the vacuum and the world with 2 light flavours and that with 3 are not the same [23].

Since scalars are so intimately tied to the structure of the QCD vacuum, their nature is something we need to understand. Here we will first focus on the $f_0(980)$ and $a_0(980)$. It has been proposed for decades that these states have one of three possible compositions: either a simple $q\bar{q}$ structure (which for the $f_0(980)$ is dominated by an $s\bar{s}$ component, since we know it couples strongly to $K\bar{K}$ ), or a tightly bound four quark system or a looser $K\bar{K}$ molecule.

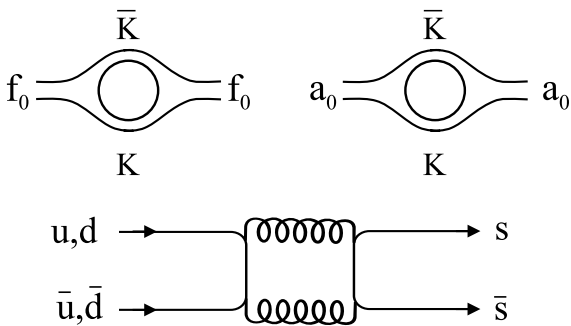

**Figure 3**. The observed properties of the two scalar mesons $f_0(980)$ and $a_0(980)$ are produced by "dressing". These states then enhance the coupling of $u\bar{u}$, $d\bar{d}$ systems to $s\bar{s}$ with no OZI suppression in scalar channels.

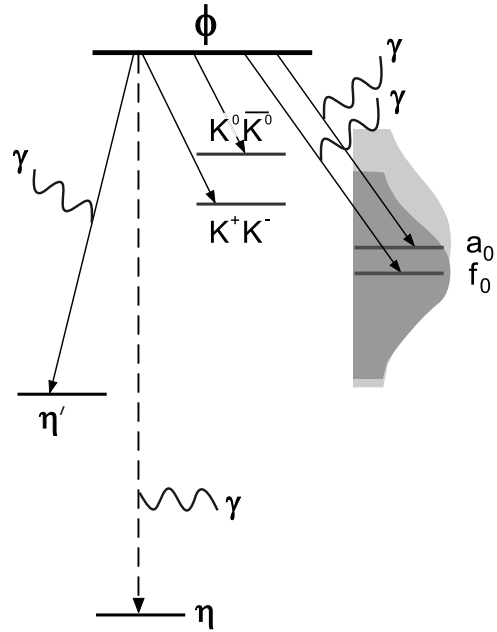

**Figure 4**. Decays of the $\phi$-meson to states nearby in energy.

The fact that the $\phi$ has a known $s\bar{s}$ composition has long made it a favourite signature for flavour tagging. $e^+e^-$ colliders running at the $\phi$ mass, like VEPP-2M at Novosibirsk and DAΦNE at Frascati, provide copious events on all $\phi$ decays, like those shown in Fig. 4. They are thus an ideal source for precise information on $\phi$-radiative decays, Figs. 4,5. Indeed it is believed that having a known initial state makes such decays a key way to distinguish between the different options for the composition of the $f_0$ and $a_0$. It was long ago that Achasov [24], and then Close and Isgur [25], and others [26] advertised that these different compositions give rise to quite distinctive branching ratios, which for $\phi \to \gamma f_0(980)$ are given in Table 1. There are analogous predictions for the $a_0(980)$: for instance, in the $K\bar{K}$ molecule picture, where $K^+K^-$ loops are key, the ratio $BR(\phi \to \gamma a_0)/BR(\phi \to \gamma f_0)$ would clearly be one if the $a_0$ and $f_0$ were degenerate in mass. The real world is more complicated as we shall see, and the models used to predict the branching ratios for the 3 options shown in Table 1 are likely too simplistic.

| Composition | BR($\phi \to \gamma f_0(980)$) |
|---|---|
| $qq\bar{q}\bar{q}$ | $O(10^{-4})$ |
| $s\bar{s}$ | $O(10^{-5})$ |
| $K\bar{K}$ | $< O(10^{-5})$ |

**Table 1**. Predictions for the absolute rate for $\phi \to \gamma f_0(980)$ depending on the composition of the $f_0(980)$.

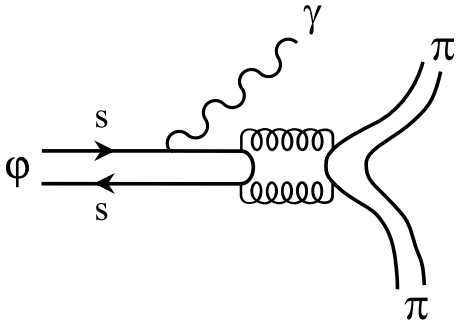

**Figure 5**. Quark line diagram for $\phi$-radiative decay to $\pi\pi$, indicating the $s\bar{s}$ to light quark transition displayed in Fig. 3.

$\phi \to \gamma\pi\pi$ and $\phi \to \gamma\pi\eta$ have recently been measured at VEPP-2M with both the SND [27] and CMD-2 [28] detectors and at DAΦNE in the KLOE experiment [29,30]. The $\pi\pi$ and $\pi\eta$ spectra, Figs. 6 and 7, show a peaking at the end of phase space that the experiments identify with the $f_0(980)$ and $a_0(980)$, respectively.

Let us look at how the KLOE group [30] use the data of Figs. 6,7 to determine the decay rate of the $f_0(980)$ to compare with the predictions of Table 1. First they fit a Breit-Wigner form to the shape of the distribution, by adjusting the mass and width of the $f_0$. One finds that this fails to describe the distribution at low $\pi\pi$ masses, so part of the decay system must be provided by something other than the $f_0(980)$. There is, of course, the background to the same $\gamma\pi\pi$ final state from the sequential decays of $\phi \to \rho\pi$ and $\rho \to \gamma\pi$. In Fig. 6 is shown the Dalitz plot for the $\gamma\pi\pi$ final state. The $\rho$ bands in the two $\gamma\pi$ channels are drawn. One sees that they have little effect on the higher $\pi\pi$ mass region that is dominated by peaking towards the end of phase-space. With the higher statistics from DAPHNE, the KLOE collaboration can separate out this $\rho\pi$ decay by its distinctive angular distribution. One thus knows for the KLOE results that the $\gamma\pi\pi$ distribution (Fig. 7) is controlled by $S$-wave $\pi\pi$ interactions. As we have seen these cannot just come from $f_0(980)$ production. The remainder is regarded as the effect of the $\sigma$. The $\sigma$'s parameters are then taken from the Fermilab E791 experiment [31] (an experiment we will refer to again later). This gives a mass of 478 MeV and a width of 324 MeV. The amplitudes for $\sigma$ and $f_0(980)$ production are then simply added and allowed to interfere. The resulting fit gives BR$(\phi \to \gamma f_0(980)) = (4.4 \pm 0.4) \cdot 10^{-4}$ and one concludes the $f_0(980)$ is a four quark system — see Table 1. A similar analysis [29] of the $\eta\pi$ channel gives a branching ratio for the $a_0(980)$ in $\phi$-radiative decay a factor of 6 smaller. So why should we redo this analysis?

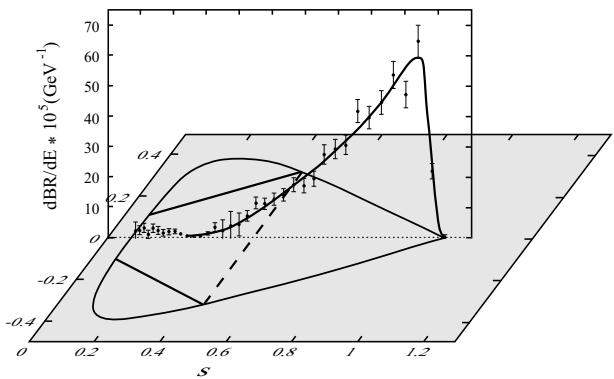

**Figure 6**. On the grey horizontal plane the Dalitz plot for $\phi \to \gamma\pi\pi$ is marked with $s$ equal to the square of the dipion invariant mass. The straight solid lines mark the central position of the $\rho$-contribution in the two $\gamma\pi$ channels. In the vertical plane is the projection of the decay distribution on the $s$-axis.

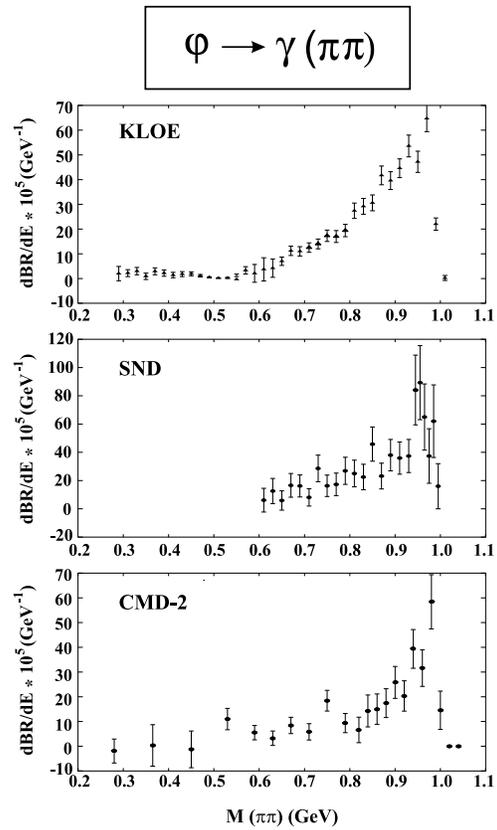

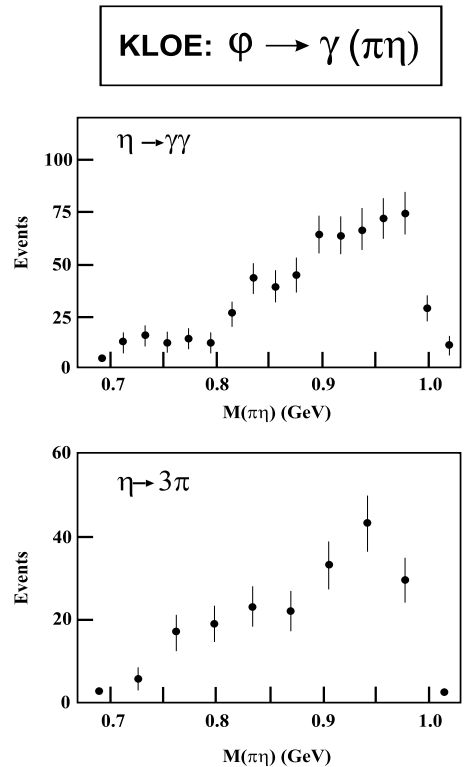

**Figure 7**. $\pi^0\pi^0$ and $\pi^0\eta$ mass distributions in $\phi$-radiative decays. The $\pi\pi$ results are from the KLOE [30], SND [27] and CMD-2 [28] detectors. The $\pi\eta$ data displayed are from KLOE [29] in both the $\gamma\gamma$ and $3\pi$ decay modes of the $\eta$.

Let us first concentrate on the isoscalar channel. The parameters of the $f_0(980)$ are not free variables. The same resonance pole position must appear in all processes to which the state couples. That is the very essence of what defines the existence of a state. Consequently, one cannot permit the mass to be reduced by 10 MeV or more, or the width of a state that is typically found to be 40-60 MeV wide cannot be allowed to be 200-250 MeV wide. These changes made by KLOE increase the branching ratio through the $f_0(980)$ by an order of magnitude. To see why, consider the key dynamics of the process. The phase space involves a product of the $\pi\pi$ momentum and the photon momentum. But because the photon is not any massless particle, but couples through a conserved current, the decay distribution in fact involves the cube of the photon momentum, as Achasov [32] has repeatedly emphasised. In terms of invariants this is proportional to $(m_\phi^2 - M(\pi\pi)^2)^3$. With the $f_0(980)$ so close to the end of phase space, small changes in its mass dramatically alter its branching ratio. This experiment cannot determine the parameters of resonances on its own. These parameters must however be the same as those required to describe other data. Moreover, the contribution of the $\sigma$ and the $f_0(980)$ must be added in a way consistent with unitarity. How to do this has recently been worked out by Elena Boglione and myself [33].

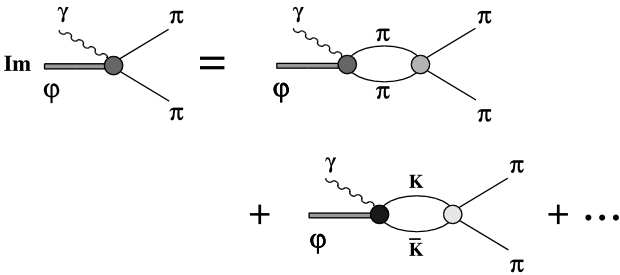

**Figure 8**. The unitarity constraint relevant to $\pi\pi$ interactions in a definite partial wave in $\phi$-radiative decay assuming the dominance of final state di-meson interactions.

To proceed the basic assumption is that any $\phi\pi$ interactions are small and the dominant strong interaction is between the final state pions. This is the presumption implicit in any isobar modelling of a decay. With the upper mass range fixed by the $\phi$ mass, there are very limited ways a $\pi\pi$ final state can be produced, Fig. 8. Either the $\phi$ radiates a photon leaving a $\pi\pi$ system that then interacts, or the $\phi$ radiates a photon producing a $K\overline{K}$ system that then interacts to produce a dipion pair. This occurs with the $\pi^0\pi^0$ system being in an isoscalar state, which is predominantly $S$-wave. With these assumptions, coupled channel unitarity requires that the amplitude, $F$ for $\phi \to \gamma\pi\pi$ can be related to the basic amplitudes $T$ for $\pi\pi \to \pi\pi$ ($T_{11}$) and $\overline{K}K \to \pi\pi$ ($T_{21}$) by

$$F(\phi \to \gamma\pi\pi) = \alpha_1(s) \cdot T(\pi\pi \to \pi\pi)$$
$$+ \alpha_2(s) \cdot T(\overline{K}K \to \pi\pi), \quad (1)$$

where $s = M(\pi\pi)^2$ and the two functions $\alpha_i(s)$ are real. They can be interpreted as the intrinsic couplings for $\phi \to \gamma\pi\pi$ and $\phi \to \gamma\overline{K}K$, respectively for $i = 1, 2$ (Fig. 8). Since the whole point of studying this process is because the initial state is almost 100% $s\overline{s}$, we expect the coupling $\alpha_2$ to be much larger than $\alpha_1$. Our analysis shows that experiment does indeed support this. However, the functions $\alpha_i(s)$ each have a factor of $(m_\phi^2 - s)$, as previously explained, following from QED gauge invariance. This strongly reduces the contribution from $T(\overline{K}K \to \pi\pi)$, which is dominated by the $f_0(980)$, and enhances the contribution from $T(\pi\pi \to \pi\pi)$, which is controlled by the $f_0(400 - 1200)$ (or $\sigma$), and from which the $f_0(980)$ effectively decouples — see Fig. 9. So while there is a sizeable $f_0(980)$ component, much of the $\pi\pi$ decay distribution is produced by $\pi\pi$ interactions outside the narrow $f_0(980)$ region.

By building into the analysis known experimental information on the scattering reactions $\pi\pi \to \pi\pi$ and $\pi\pi \to \overline{K}K$, which automatically embodies details of the $f_0(400-1200)$ and $f_0(980)$, we can use the data on $\phi \to \gamma\pi^0\pi^0$ to determine the couplings of these scalar resonances to this channel in as model-independent a way as possible. This is the purpose of the recent analysis by Elena Boglione and myself [33]. As a simple template we first used the old hadronic amplitudes determined by David Morgan and I [34] (called ReVAMP as explained in Ref. 33). These have the $f_0$-pole on sheet II at $\sqrt{s} = (988 - i \cdot 23)$ MeV. Factoring out the Adler zero and the photon momentum required by QED gauge invariance for the radiative decay process, we then have constant coupling functions, $\alpha_i(s)$, and obtain the fit shown in Fig. 10. The quality of the fit is excellent indicating no reason to sacrifice our starting assumptions and showing the final state $\pi\pi$ interactions in this decay are completely consistent with those from other processes built into the ReVAMP amplitudes.

It should be clear that the exact branching fraction for the $f_0(980)$ is exceedingly sensitive to the position of the corresponding pole, because of its nearness to the edge of phase space. Moreover, one does not see the $f_0(980)$ as a simple Breit-Wigner shape (compare Fig. 9b with Fig. 10), so in fact its branching fraction is not directly related to an experimental observable. The only well-defined quantity is the residue at the pole which truly represents the coupling $\phi \to \gamma f_0$. For a state that overlaps with another broad resonance and with a strongly coupled threshold, the idea of a branching fraction is wholly model-dependent.

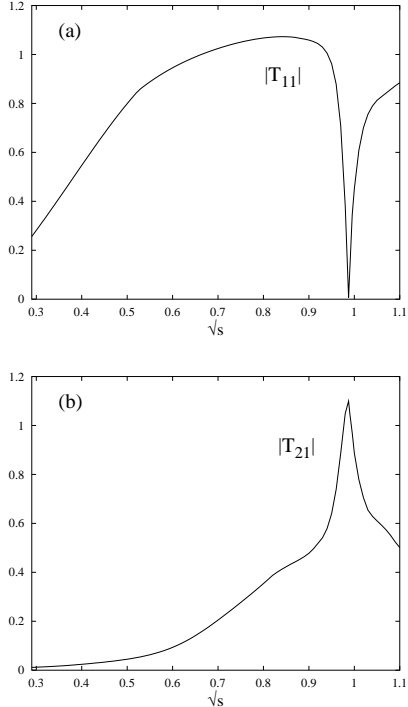

**Figure 9**. Plots (**a**) and (**b**) show the $I = J = 0$ ReVAMP hadronic amplitudes [34] $|T(\pi\pi \to \pi\pi)|$ (labelled $T_{11}$) and $|T(\overline{K}K \to \pi\pi)|$ (labelled $T_{21}$) from which $F(s)$ is constructed, according to Eq. (1), where $\sqrt{s} = M(\pi\pi)$, the c.m. energy.

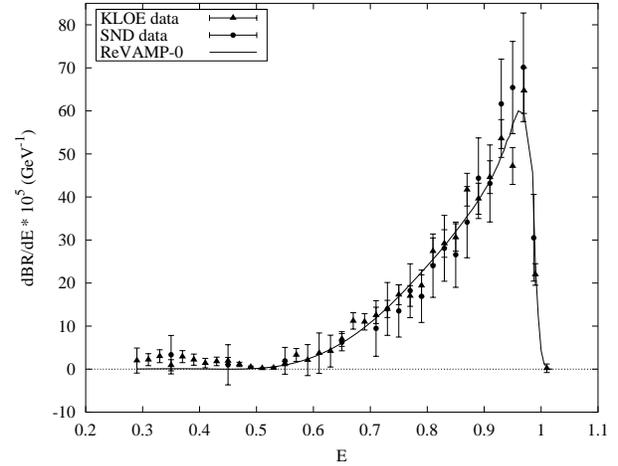

**Figure 10**. Simultaneous fits to the data on the $\pi^0\pi^0$ decay distribution in $\phi \to \gamma\pi^0\pi^0$ from the KLOE [30] and SND [27] collaborations. These fits have been obtained using the ReVAMP set of underlying amplitudes, with just 3 real parameters [33], as described in the text. $E = M(\pi\pi)$.

The fit shown in Fig. 10 gives a $\phi \to \gamma f_0(980)$ coupling which approximates to 10% of the total $\pi\pi$ S-wave decay being through the $f_0(980)$, i.e. a branching fraction of $0.31 \cdot 10^{-4}$. Looking at Table 1 we see this corresponds no longer to the $qq\overline{qq}$ composition inferred by KLOE [30], but being an order of magnitude smaller is like that for an $s\overline{s}$ constitution. However, this is too simplistic. Oller [35] has noted that the $K\overline{K}$ molecular picture can include not just the charged kaon loops used in the predictions of Table 1, but also neutral kaons too. Then the prediction extends from $10^{-6}$ to $10^{-4}$ to encompass our result too.

What of $\phi \to \gamma a_0(980) \to \gamma(\pi^0\eta)$? A coupled channel analysis like that for the $\pi^0\pi^0$ decay mode is not presently practicable because of lack of precision information on $\pi\eta \to \pi\eta$ and $\pi\eta \to \overline{K}K$ scattering in S-wave channels. However, these have none of the complication of overlapping resonances of the isoscalar $\pi\pi$ mode. Consequently, it is more likely that such an analysis when possible will reveal a $BR(\phi \to \gamma a_0(980))$ much more similar to that found by KLOE [29]. This is the same order of magnitude as our smaller $BR(\phi \to \gamma f_0(980))$, which is something expected if there is a large $K\overline{K}$ component of these two scalar mesons. Of course, the most simple minded interpretation of the $K\overline{K}$ molecule picture with just charged kaon loops seeding the decay gives the ratio of branching ratios for $a_0$ and $f_0$ to be one, only if these states are degenerate in mass. Just a 10 MeV difference in their masses changes this to 0.4 or 2, depending on the sign of the mass difference. Consequently, model predictions and analyses of data are extremely sensitive to the $f_0(980)$ and $a_0(980)$ pole positions.

Elena Boglione and I [33] have illustrated this by considering in our analysis of the $\pi^0\pi^0$ data the more recently determined hadronic amplitudes of Anisovich and Sarantsev [36]. These give the $f_0$-pole at $\sqrt{s} = (1024 - i \cdot 43)$ MeV. Fits displayed in Ref. 33 show that this much greater width for the $f_0(980)$ results in a coupling, which translates into a modelled branching ratio of $1.9 \cdot 10^{-4}$ — a factor 6 larger than using the ReVAMP amplitudes with their much narrower $f_0$ we have just described. However, the use of Anisovich and Sarantsev amplitudes (AS) gives fits of much poorer quality. If it were not for the fact that these authors treat a far greater range of more recent data, such as that from Crystal Barrel and from GAMS (see Ref. 36 for details), than in the older ReVAMP analysis [34], the quality of fit would dismiss such a wide $f_0(980)$ as unlikely. Nevertheless, the fact that the $\phi \to \gamma f_0$ coupling is so sensitive to the details of the $f_0$-pole means we must tighten up our determination of its position.

States in the spectrum are identified not by peaks in cross-sections, but as poles of the $S$-matrix. Crucially, their position in the complex energy plane is independent of the process in which they appear. Precision knowledge of the pole positions and residues of the $f_0/a_0(980)$ is essential if we are to draw conclusions about the composition of these states in as model-independent way as possible — independently of the modelling of $\phi$-radiative decay presented here or that involving $K\bar{K}$ loops by Achasov [24] or the generation of these states by unitarisations of chiral dynamics [37,35] — just using experiment. A rich source of additional information on light hadron final states is being provided by the decays of heavy flavour mesons. Analysis of the Dalitz plot distribution of new results on $D_s \to 3\pi$ and on $J/\psi \to \phi\pi\pi$, have the capability to fix the $f_0(980)$ parameters we need: studies are under way.

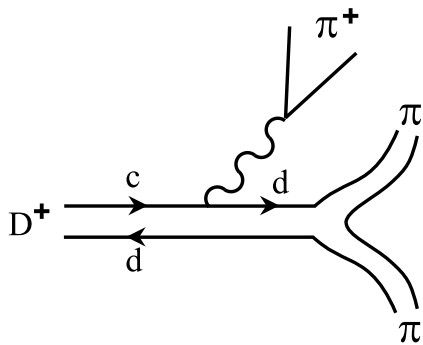

**Figure 11**. Quark line graph for the weak decay $D^+ \to \pi^+\pi^+\pi^-$.

An important illustration of how final state interactions and their universality shape such decays has recently been provided by studies of $D \to 3\pi$ [38]. Decay data are typically analysed in an isobar picture [31,38], where it is assumed the three final particles interact only in pairs with the third as a spectator, Fig. 11. If 3-body forces are needed, they are usually assumed to have constant matrix elements and so populate the Dalitz plot according to phase-space. In the analysis of the Fermilab E791 results [31] with $\sim 1100$ $D^+ \to \pi^+\pi^+\pi^-$ events, the known resonances that couple to $\pi\pi$ — the $\rho$, $f_2(1270)$, etc. — are included. The resulting fit is poor at low $\pi\pi$ masses, cf. Fig. 12a. However, this is dramatically improved if an $I = J = 0$ resonance of mass $(478 \pm 24 \pm 17)$ MeV and width $(324 \pm 42 \pm 21)$ MeV is added in Breit-Wigner form, cf. Fig. 12b. Hence the E791 group claim to have confirmed the $\sigma$ resonance [31]. This is to forget that, assuming an isobar model, one has by definition now determined the $I = J = 0$ $\pi\pi \to \pi\pi$ interaction. Though the phase-shift is large, it is not given by a simple Breit-Wigner with the claimed mass and width alone.

With $\sim 1500$ events on this same decay, the FOCUS group [38] at Fermilab has confirmed the E791 data and analysis, Fig. 12. But if instead the low mass $S$-wave $\pi\pi$ interaction is parametrised using the Anisovich and Sarantsev description, then the fit to the $D \to 3\pi$ Dalitz plot is even better, Fig. 12c. Though here and in Ref. 33, we have queried the wide $f_0(980)$ of AS, this matters little for $D \to 3\pi$ decay. Their amplitudes and those of ReVAMP are virtually the same from $\pi\pi$ threshold to 900 MeV (see Fig. 1 of Ref. 33). What Malvezzi and collaborators [38] have shown is that the $\pi\pi$ final state interactions in $D \to 3\pi$ decay are completely consistent with what we know of such interactions from all other processes.

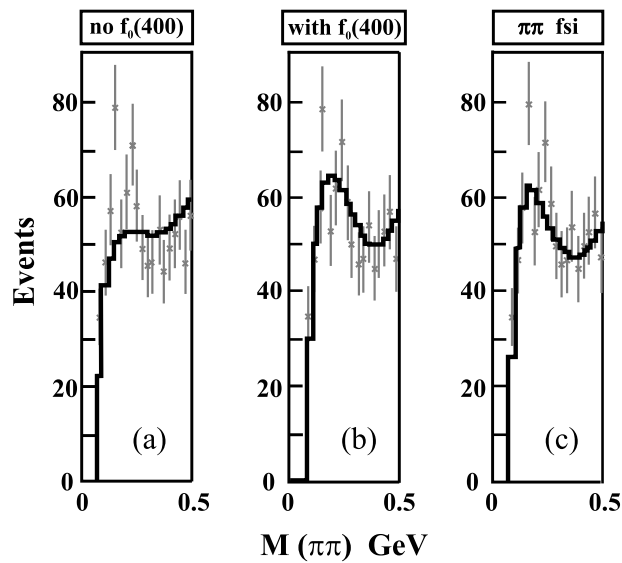

**Figure 12**. Low $\pi\pi$ mass distribution in $D^+ \to \pi^+\pi^+\pi^-$ decay from the FOCUS group [38]. (a) shows the result of a fit to the full Dalitz plot with no $f_0(400-1200)$, (b) the improvement when scalar $\pi\pi$ interactions are represented by a Breit-Wigner for an $f_0(400)$, (c) the even greater improvement when the final state interactions (fsi) are those of experimental $\pi\pi$ scattering [38].

Whether there is a $\sigma$ or not as a short-lived resonance is then not a question of whether it is just seen in $D \to 3\pi$ decay, but whether it is there in all other $I = J = 0$ $\pi\pi$ final states too. The difficulty in answering the question of whether data on the real axis have sufficient precision to determine the existence of such a very distant pole [39] is amply illustrated by the difference between the AS [36] and ReVAMP [34] amplitudes. AS have no low mass pole, while ReVAMP does : yet each describes essentially the same experimental results below 900 MeV.

What we learn is that analysing data on a single channel, where final state interactions are important, cannot be meaningfully done in isolation. Unitarity requires consistency between reactions. Only by analysing data from different processes with the same final states simultaneously can we hope to be able to draw definitive conclusions about the fascinating scalar sector. Since these states with zero quantum numbers reflect the nature of the QCD vacuum, they will remain in the spotlight for some time to come.

## Acknowledgments

It is pleasure to thank Guiseppe Nardulli and his team of organisers, particularly Pietro Colangelo and Fulvia De Fazio, for inviting me to give this presentation amongst the artistic collection of Castello Aragonese and for their very warm hospitality. I acknowledge the partial support of the EU-RTN Programme, Contract No. HPRN-CT-2002-00311, "EURIDICE" for this work.

## References

1. see, for instance, M.R. Pennington, "Pions: unbaring their lightness of being", Proc. 14th Summer School *Understanding the Structure of Hadrons*, Prague, 2001, Czech. J. Phys. **52(B)** (2002) 28.
2. see, for instance, M.R. Pennington, "Translating quark dynamics into hadron physics (and back again)," Proc. *7th Int. Workshop on Meson Production, Properties and Interaction (Meson 2002)*, Cracow ( ed. L. Jarczyk *et al.*) (pub. World Scientific, 2003) pp. 1-14 [hep-ph/0207220].
3. M. A. Shifman, A. I. Vainshtein and V. I. Zakharov, Nucl. Phys. **B147** (1979) 385, 448.
4. P. Colangelo and A. Khodjamirian, [hep-ph/0010175]; F. De Fazio, Boris Ioffe Festschrift *At the Frontier of Particle Physics Handbook of QCD*, ed. M. Shifman, (World Scientific), pp. 1671-1717 [hep-ph/0010007].
5. M. R. Pennington, Nucl. Phys. **A623** (1997) 189C [hep-ph/9612417].
6. J. Stern, H. Sazdjian and N.H. Fuchs, Phys. Rev. **D47** (1993) 3814; M. Knecht and J. Stern, *Second DA$\Phi$NE Physics Handbook*, (ed. L. Maiani, G. Pancheri and N. Paver) (pub. INFN, Frascati, 1995) pp. 169-190.
7. G. Colangelo, J. Gasser and H. Leutwyler, Nucl. Phys. **B603** (2001) 125 [hep-ph/0103088].
8. S. Adler, Phys. Rev. **137** (1965) B1022, *ibid.* **139** (1965) B1638.
9. J. Gasser and H. Leutwyler, Ann. Phys. (NY) **158** (1984) 142.
10. L. Afanasev *et al.* [DIRAC experiment], Proc. *9th Int. Conf. on Hadron Spectroscopy (Hadron 2001)*, Protvino (ed. D. Amelin and A. M. Zaitsev) (pub. Amer. Inst. Phys.) AIP Conf. Proc. **619** (2002) 745.
11. S. Pislak *et al.* [BNL-E865], Phys. Rev. Lett. **87** (2001) 221801.
12. G. Colangelo, J. Gasser and H. Leutwyler, Phys. Rev. Lett. **86** (2001) 5008 [hep-ph/0103063].
13. C. D. Roberts and A. G. Williams, Prog. Part. Nucl. Phys. **33** (1994) 477 [hep-ph/9403224]; P. Maris and C. D. Roberts, Int. J. Mod. Phys. E **12** (2003) 297 [nucl-th/0301049].
14. R. Alkofer and L. von Smekal, Phys. Rept. **353** (2001) 281 [hep-ph/0007355], Nucl. Phys. **A680** (2000) 133 [hep-ph/0004141].
15. C. S. Fischer and R. Alkofer, Phys. Rev. **D67** (2003) 094020 [hep-ph/0301094].
16. K.Langfeld *et al.*, Phys. Rev. **C67** (2003) 065206.
17. P. Maris and C. D. Roberts, Phys. Rev. **C56** (1997) 3369 [nucl-th/9708029].
18. P. Maris, C. D. Roberts and P. C. Tandy, Phys. Lett. **B420** (1998) 267 [nucl-th/9707003].
19. F. D. Bonnet *et al.*, Phys. Rev. **D62** (2000) 051501 [hep-lat/0002020].
20. K. Benhaddou, P. Watson and M.R. Pennington, *in preparation*.
21. E. Van Beveren *et al.*, Z. Phys. **C30** (1986) 615; N. A. Tornqvist, Z. Phys. **C68** (1995) 647; M. Boglione and M. R. Pennington, Phys. Rev. Lett. **79** (1997) 1998 [hep-ph/9703257].
22. P. Geiger and N. Isgur, Phys. Rev. **D47** (1993) 5050.
23. S. Descotes and J. Stern, Phys. Lett. **B488** (2000) 274 [hep-ph/0007082]; S. Descotes, JHEP **0103** (2001) 002 [hep-ph/0012221].
24. N.N. Achasov and V.N. Ivanchenko, Nucl. Phys. **B315**, 465 (1989); N.N. Achasov and V.V. Gubin, Phys. Rev. **D63**, 094007 (2001); N.N. Achasov, Proc. *9th Int. Conf. on Hadron Spectroscopy (Hadron 2001)*, Protvino (ed. D. Amelin and A.M. Zaitsev) (pub. Amer. Inst. Phys.) AIP Conf. Proc. **619** (2002) pp. 112-121 [hep-ph/0110059].
25. F. E. Close, N. Isgur and S. Kumano, Nucl. Phys. **389** (1993) 513 [hep-ph/9301253].
26. J.Lucio and J. Pestieau, Phys. Rev. **D42** (1990) 3253; S. Nussinov and T.N. Truong, Phys. Rev. Lett. **63** (1989) 2003; N. Paver and Riazuddin, Phys. Lett. **B246** (1990) 240.
27. M. N. Achasov *et al.* [SND], Phys. Atom. Nucl. **62** (1999) 442 [Yad. Fiz. **62** (1999) 484], Phys. Lett. **B485** (2000) 349.
28. R.R. Akhmetshin *et al.* [CMD-2], Phys. Lett. **B462** (1999) 380.
29. A. Aloisio *et al.* [KLOE], Phys. Lett. **B536** (2002) 209 [hep-ex/0204012].
30. A. Aloisio *et al.* [KLOE], Phys. Lett. **B537** (2002) 21 [hep-ex/0204013].
31. E.M. Aitala *et al.* [Fermilab E791], Phys. Rev. Lett. **86** (2001) 770 [hep-ex/0007028].
32. e.g. N.N. Achasov, *Second DA$\Phi$NE Physics Handbook*, (ed. L. Maiani, G. Pancheri and N. Paver) (pub. INFN, Frascati, 1995) pp. 671-680.
33. M. Boglione and M. R. Pennington, EPJC *in press* [hep-ph/0303200].
34. D. Morgan and M. R. Pennington, Phys. Rev. **D48** (1993) 1185.
35. J. A. Oller, Nucl. Phys. **A714** (2003) 161 [hep-ph/0205121].
36. V. V. Anisovich and A. V. Sarantsev, Eur. Phys. J. **A16** (2003) 229 [hep-ph/0204328].
37. E. Marco, S. Hirenzaki, E. Oset and H. Toki, Phys. Lett. **B470** (1999) 20 [hep-ph/9903217].
38. S. Malvezzi, "Light quark and charm interplay in the Dalitz-plot analysis of hadronic decays in FOCUS," [hep-ex/0307055].
39. M.R. Pennington, "Riddle of the scalars: where is the $\sigma$?", Proc. *Int. Workshop on Hadron Spectroscopy*, Frascati, 1999 (ed. T. Bressani *et al.*) (pub. Frascati Physics Series, Vol. 15) pp. 95-114 [hep-ph/9905241].